# The thermodynamic stability and hydration enthalpy of strontium cerate doped by yttrium

N.I. Matskevich *(Nikolaev Institute of Inorganic Chemistry, Siberian Branch of the Russian Academy of Sciences, Novosibirsk, Russia),* Th. Wolf *(Forschungszentrum Karlsruhe, Institute of Solid State Physics, Karlsruhe, Germany),* M.Yu. Matskevich *(Nikolaev Institute of Inorganic Chemistry, Siberian Branch of the Russian Academy of Sciences, Novosibirsk, Russia)*

**Abstract**

The standard molar enthalpy of formation of $SrY_{0.05}Ce_{0.95}O_{2.975}$ has been derived by combining the enthalpy of solution of this compound in 1 M HCl + 0.1 KI obtained by us and auxiliary literature data. The following value has been derived: $\Delta_f H^o(SrY_{0.05}Ce_{0.95}O_{2.975}$, s, 298.15 K) = −1720.4 ± 3.4 kJ/mol. The obtained value has been used to obtain the formation enthalpy of $SrY_{0.05}Ce_{0.95}O_{2.975}$ from the mixture of binary oxides ($\Delta_{ox}H^0$(298.15 K) = –45.9 ± 3.4 kJ/mol) and formation enthalpy of reaction of $SrY_{0.05}Ce_{0.95}O_{2.975}$ with water forming $Sr(OH)_2$, $CeO_2$, $Y_2O_3$ ($\Delta_r H^0$ (298.15 K) = –85.5 ± 3.4 kJ/mol). Data obtained by solution calorimetry and additional information on the entropies of different substances have shown that $SrY_{0.05}Ce_{0.95}O_{2.975}$ is thermodynamically stable with respect to a mixture of SrO, $Y_2O_3$, $CeO_2$ and that the reaction of $SrY_{0.05}Ce_{0.95}O_{2.975}$ with water is thermodynamically favourable.

## 1. Introduction

Proton conductors on the basis of alkali-earth (barium or strontium) cerates are of considerable interest for application in hydrogen sensors, electrocatalytic reactors,

---

*Corresponding author. Tel.: +7–3833–306449; fax: +7–3833–309489

E–mail address: nata@che.nsk.su
steam electrolyzers, and as electrolytes in fuel cells [1]. The substituted solid solutions based on the perovskite oxides $MCeO_3$ (M = Ba, Sr), in which RE replaces Ce, is a typical example of this class of materials. The general formula is written as $BaCe_{1-x}RE_xO_{3-x/2}$, where RE is a rare earth element.

As it is well known [1, 2], solid oxide fuel cells typically operate at a temperature of 1273 K. Doped barium cerates achieve this target at temperature around 873 K. Operation at lower temperature potentially offers reduced materials and process costs and increased thermodynamic efficiency, provided the necessary solid electrolyte and electrodes can be developed. There is considerable emphasis on moving to lower temperature of operation in the next generation of fuel cells. A highly attractive alternative may be to use proton conducting electrolytes such as the perovskite $BaCeO_3$ (or $SrCeO_3$) instead of doped zirconium oxide to provide good ionic conductivity in the temperature range of 500–750 °C. In order to use these alternative electrolytes systems, it is necessary to perform detail physico-chemical study of employed compounds. There are no data on thermodynamics of $SrCe_{1-x}Y_xO_{3-x/2}$ solid solutions in literature.

The objective of this work was to measure formation enthalpy of $SrCe_{1-x}Y_xO_{3-x/2}$ and to study their thermodynamic stability and hydration enthalpy.

## 2. Experimental

$SrY_{0.05}Ce_{0.95}O_{2.975}$(s) was prepared by solid state synthesis heating stoichiometric amounts of dried $CeO_2$ (s), $SrCO_3$ (s), $Y_2O_3$ (s) (Cerac, mass fractions of compounds are more than 0.9999). Stoichiometric mixtures of $SrCO_3$, $CeO_2$, $Y_2O_3$ were mixed and milled in a planetary mill with agate balls during 12 h with intermediate regrinding. Then the powders were pressed into pellets with a diameter of 10 mm and calcinated at 1000 °C during 70 h. X-ray analysis showed that these conditions were not sufficient to prepare phase pure $SrCe_{0.95}Y_{0.05}O_{2.975}$. The samples consisted of a mixture of $CeO_2$ and $SrCe_{0.95}Y_{0.05}O_{2.975}$. Therefore after regrinding a second sintering was carried out at 1300° C during 17 h. A final sintering at 1500 °C during 40 h resulted in a phase pure $SrCe_{0.95}Y_{0.05}O_{2.975}$ sample. The compound was characterized by X–ray powder diffraction and chemical analysis [3]. According to the results of the analyses the involved compound was found to be single phase with an accuracy of about 1%.

Solution calorimetry was used as an investigation method. The experiments were performed in an automatic dissolution calorimeter with an isothermal shield. The construction of the solution calorimeter and the experimental procedure are described elsewhere [4–5]. The volume of the calorimetric vessel was 200 ml. The reproducibility of the heat equivalent of the calorimeter with an automatic calibration system was 0.03%. To check the precision of the calorimeter, dissolution of a standard substance, potassium chloride, was performed. The obtained dissolution heat of KC1 (17.529+0.009 kJ/mol) was found to be in a good agreement with the value recommended in the literature (17.524+0.007 kJ/mol) [6]. The amounts of substances used were 0.05—0.1 g.

Calorimetric cycles were designed in such a way that it was possible to determine the formation enthalpy of $SrY_{0.05}Ce_{0.95}O_{2.975}$ on the basis of measured and literature reference data. The solution of 1 M HCl with 0.1 M KI was using as solvent. KI was added to account for the reduction of $Ce^{4+}$ to $Ce^{3+}$. The thermochemical cycle, from which the enthalpy of formation of $SrY_{0.05}Ce_{0.95}O_{2.975}$ was derived, was given in Table 1.

Table 1. Reaction scheme to perform the standard molar enthalpy of formation of $SrY_{0.05}Ce_{0.95}O_{2.975}$ at the temperature of 298.15 K

| Nr. | Reaction | $\Delta_{sol}H_m^o$ (kJ · mol$^{-1}$) | Ref. |
|---|---|---|---|
| 1 | $SrY_{0.05}Ce_{0.95}O_{2.975}(s) + (5.95\ HCl + 0.95\ KI)_{sol} = (SrCl_2 + 0.05YCl_3 + 0.95CeCl_3 + 0.95KCl + 0.475I_2 + 2.975\ H_2O)_{(sol)}$ | –323.9 ± 3.1 | this work |
| 2 | $SrCl_2(s) + 0.05YCl_3(s) + 0.95CeCl_2(s)$ + solution 1 = $(SrCl_2(s) + 0.05YCl_3(s) + 0.95CeCl_2)_{(sol)}$ | –181.32 ± 0.55 | [7] |
| 3 | $2.975H_2(g) + 1.4875O_2(g)$ + solution 1 = $2.975H_2O_{(sol)}$ | –850.37 ± 0.13 | [8] |
| 4 | $0.95KI(s)$ + solution 1 = $0.95KI_{(sol)}$ | +19.79 ± 0.33 | [7] |
| 5 | $0.95K(s) + 0.475I_2(s) = 0.95KI_{(sol)}$ | –312.69 ± 0.13 | [9] |
| 6 | $0.475I_2(s)$ + solution 1 = $0.475I_{2(sol)}$ | +2.65 ± 0.48 | [7] |
| 7 | $0.95KCl(s)$ + solution 1 = $0.95KCl_{(sol)}$ | +17.11 ± 0.05 | [9] |
| 8 | $0.95K(s) + 0.475Cl_2(g) = 0.95KCl(s)$ | –414.64 ± 0.12 | [9] |
| 9 | $2.975H_2(g) + 2.975Cl_2(g)$ + solution 1 = $5.95HCl_{(sol)}$ | –977.94 ± 0.06 | [8] |
| 10 | $Sr(s) + Cl_2(g) = SrCl_2(s)$ | –832.43 ± 0.85 | [10] |
| 11 | $0.95Ce(s) + 1.425Cl_2(g) = 0.95CeCl_3(s)$ | –1007.51 ± 0.50 | [11] |
| 12 | $0.05Y(s) + 0.075Cl_2(g) = 0.05YCl_3(s)$ | –48.68 ± 0.15 | [11] |
| 13 | $Sr(s) + 0.05Y(s) + 0.95Ce(s) + 1.4875O_2(g) = SrY_{0.05}Ce_{0.95}O_{2.975}(s)$ | –1720.45 ± 3.36 | this work |

Here: solution 1 is a solution of 1 M HCl with 0.1 M KI; s – solid; g – gas; $\Delta_{sol}H_m^o$ – the molar enthalpy of solution;

## 3. Results

We measured only the dissolution of $SrY_{0.05}Ce_{0.95}O_{2.975}$ in 1 M HCl with 0.1 M KI. $\Delta_{sol}H^0(SrY_{0.05}Ce_{0.95}O_{2.975}$, 298.15 K) = –323.9 ± 3.1 kJ/mol. The dissolution enthalpies were calculated as average values of six calorimetric experiments. Errors were calculated for the 95% confidence interval using the Students coefficient.

Other solution enthalpies, which were necessary to calculate the enthalpy of formation of $SrY_{0.05}Ce_{0.95}O_3$, were taken from literature references (see, Table 1). The following equation can be written to calculate the formation enthalpy of $SrY_{0.05}Ce_{0.95}O_{2.975}$:

$$Sr(s) + 0.95Ce(s) + 0.05Y(s) + 1.4875O_2 (g) = SrY_{0.05}Ce_{0.95}O_{2.975}(s) + \Delta_rH_{13}^o,$$

where $\Delta_rH_{13}^o = -\Delta_{sol}H_1^o + \Delta_{sol}H_2^o + \Delta_{sol}H_3^o - \Delta_{sol}H_4^o - \Delta_{sol}H_5^o + \Delta_{sol}H_6^o + \Delta_{sol}H_7^o + \Delta_{sol}H_8^o + -\Delta_{sol}H_9^o + \Delta_{sol}H_{10}^o + \Delta_{sol}H_{11}^o + \Delta_{sol}H_{12}^o$.

After combining all reactions, we obtain: $\Delta_f H^o(SrY_{0.05}Ce_{0.95}O_{2.975}$, s, 298.15 K) = −1720.45 ± 3.36 kJ/mol.

The experimental data were also used to study the thermodynamic phase stability with respect to the phase mixtures with the same nominal composition.

On the basis of formation enthalpy of $SrY_{0.05}Ce_{0.95}O_{2.975}$ obtained by us and literature data for formation enthalpies of SrO (–591.0 kJ/mol), $CeO_2$ (–1090.4 kJ/mol), $Y_2O_3$ (–1905.0 kJ/mol) [12] we calculated the enthalpies of formation of $SrY_{0.05}Ce_{0.95}O_{2.975}$ from binary oxides

$SrO(s) + 0.95CeO_2(s) + 0.025Y_2O_3(s) = SrCe_{0.95}Y_{0.05}O_{2.975}$ (s)
$\Delta_{ox}H^0$ (298.15 K) = –45.9 ± 3.4 kJ/mol

These data and entropies of all the substances employed in above reactions allow us to conclude that the formation of $SrCe_{0.95}Y_{0.05}O_{2.975}$ from the oxides is thermodynamically favourable at room temperature.

Using our experimental data for $\Delta_f H^o(SrCe_{0.95}Y_{0.05}O_{2.975}$, s), literature data for $\Delta_f H^o(CeO_2$, s), $\Delta_f H^o(Y_2O_3$, s), and formation enthalpies for $Sr(OH)_2$ (–964.3 kJ/mol), $H_2O$ (–241.9 kJ/mol) [12] it is possible to obtain thermodynamical data for the hydration enthalpy of $SrCe_{0.95}Y_{0.05}O_3$. The result is given below.

$SrY_{0.05}Ce_{0.95}O_{2.975}(s) + H_2O(g) = Sr(OH)_2(s) + 0.95CeO_2(s) + 0.025Y_2O_3(s)$
$\Delta_r H^0$ (298.15 K) = –85.5 ± 3.4 kJ/mol

The value of enthalpy of the above reaction and literature data on the entropies of $SrY_{0.05}Ce_{0.95}O_{2.975}(s)$, $H_2O(g)$, $Sr(OH)_2(s)$, $CeO_2(s)$, $Y_2O_3(s)$ allow us to conclude that the reaction of $SrY_{0.05}Ce_{0.95}O_{2.975}(s)$ with water is thermodynamically favourable. The hydration enthalpy is a very important value for the understanding of fundamental aspects of the application of electrolytes in fuel cells [1]. The high value of hydration obtained by us allows one to assume that $SrY_{0.05}Ce_{0.95}O_{2.975}$ will have high transport properties.

## 4. Conclusions

Solution calorimetry was used to measure the dissolution enthalpy of $SrY_{0.05}Ce_{0.95}O_{2.975}$ in 1 M HCl with 0.1 M KI for the first time. Based on the above value and auxiliary literature data the enthalpies of the following reactions were calculated:

1. formation enthalpy of $SrY_{0.05}Ce_{0.95}O_{2.975}$ from the mixture of Sr(s), Y(s), Ce(s), $O_2$(g) (standard formation enthalpy);
2. formation enthalpy of $SrY_{0.05}Ce_{0.95}O_{2.975}$ from the mixture of binary oxides;
3. formation enthalpy of reaction of $SrY_{0.05}Ce_{0.95}O_{2.975}$ with water forming $Sr(OH)_2$, $CeO_2$, $Y_2O_3$.

Data obtained by solution calorimetry and additional information on the entropies of different substances showed that $SrY_{0.05}Ce_{0.95}O_{2.975}$ was thermodynamically stable with respect to the mixture of SrO, $Y_2O_3$, $CeO_2$ and that the reaction with water was thermodynamically favourable.

## Acknowledgements

*This work is supported by Special program of interdisciplinary projects performed by scientists from Siberian Branch and Ural Branch (project Nr. 202) and Karlsruhe Research Center*